\documentclass[aps,pre,reprint,amsmath,amssymb,superscriptaddress]{revtex4-2}
\usepackage{color}
\usepackage[english]{babel}

\usepackage{graphicx}
\usepackage{dcolumn}
\usepackage{bm}
\usepackage{subfigure}
\usepackage{float}
\DeclareMathOperator{\sech}{sech}

\newcommand{\MST}{\langle \tilde{t} \rangle}
\newcommand{\uMST}{\langle t \rangle}

\begin{document}




\title{Active search for a reactive target in thermal environments}

\author{Byeong Guk Go}
\affiliation{Department of Physics, Korea Advanced Institute of Science and Technology, Daejeon 34141, Korea}
\author{Euijin Jeon}
\affiliation{Department of Physics, Technion–Israel Institute of Technology, Haifa 3200003, Israel}
\author{Yong Woon Kim}
\affiliation{Department of Physics, Korea Advanced Institute of Science and Technology, Daejeon 34141, Korea}

\date{\today}

\begin{abstract}
We study a stochastic process where an active particle, modeled by a one-dimensional run-and-tumble particle, searches for a target with a finite absorption strength in thermal environments. Solving the Fokker-Planck equation for a uniform initial distribution, we analytically calculate the mean searching time (MST), the time for the active particle to be finally absorbed, and show that there exists an optimal self-propulsion velocity of the active particle at which MST is minimized. As the diffusion constant increases, the optimal velocity changes from a finite value to zero, which implies that a purely diffusive Brownian motion outperforms an active motion in terms of searching time. Depending on the absorption strength of the target, the transition of the optimal velocity becomes either continuous or discontinuous, which can be understood based on the Landau approach. In addition, we obtain the phase diagram indicating the passive-efficient and the active-efficient regions. Finally, the initial condition dependence of MST is presented in limiting cases.
\end{abstract}

\maketitle

\section{Introduction}

Random target search is a problem of considerable importance in stochastic processes \cite{Redner2001, kamp92}, with applications ranging from animal foraging \cite{krapivsky1996kinetics} to recognizing a specific site by DNA-binding proteins \cite{cop04}. First-passage time is a key quantity that measures how long it takes for a searcher to arrive at a target location for the first time. Over the past decades, a number of literatures have been devoted to extending this classical subject to include various aspects, such as search by L{\`e}vy walk or flight \cite{vis99, Oben06Levy}, non-Markovian random walker \cite{guerin2016mean, levernier2022}, and the effects of external potential \cite{kusmierz2017optimal, zano21} or domain topography \cite{Condamin05, Condamin07, Benichou10Geo, Benichou14Geo, volpe17}. When there are multiple searchers, the mean first-passage times show intriguing dependence on the number or the initial distribution of searchers \cite{Kim2017, Lawley2020,  Snider2011, Kim2022}. Recently, the influence of interactions among searchers on the mean first-passage time was also investigated \cite{agranov18, ms21, ro23}.

In previous studies, an ideal target is usually assumed so that upon its first encounter, a searcher finds (or reacts with) the target with a certainty. The presence of an ideal target, as described by a perfect absorption boundary, would be rather unrealistic, especially given the possibility of error in recognition in animal foraging or a finite reaction energy in chemical reactions.  Even upon encountering, there is a finite probability for a searcher to miss a target. This kind of target is referred to as a reactive or a partially-absorbing target, which can be characterized by a radiation boundary condition in the Fokker-Planck equation \cite{Redner2001, Boccardo18, Erb07} or by assuming a stochastically gated target that switches between closed and open phases \cite{Vasquez21}.

In recent years, target search by an active matter has received significant attention \cite{Tej12, Rup16, Wang16, Zano21abp}. In contrast to passive Brownian particles, active particles exhibit self-propelled directed motions and unique non-equilibrium features in individual \cite{ramaswamy10, Mar13, Bech16} and collective behaviors \cite{solon2015, cates2015}. While several studies have been devoted to the reactive target search by a passive particle \cite{Boccardo18, Erb07}, only recently have there been a few attempts to consider an active particle searching for a reactive target. In thees attempts, the active particle was modeled by a run-and-tumble particle (RTP), and the effect of thermal diffusion was neglected \cite{Vasquez21, jeon23}. However, as shown in the active Brownian motions, active particles are also subject to thermal noises \cite{Bech16, romanczuk2012}. A comprehensive picture of the stochastic process of a reactive target search by an active particle in the presence of thermal noises is still lacking.

In this work, we address this problem by considering an RTP in a confined one-dimensional space. In particular, the presence of a reactive target and thermal diffusion are incorporated in the stochastic differential equation of an RTP. By solving the corresponding Fokker-Planck equation for a uniform initial distribution, we obtain the analytic expression of the mean searching time (MST), i.e., the time it takes for an active particle to be finally absorbed by a target, and show that there exists an optimal propulsion velocity of the active particle which minimizes MST. As the diffusion constant increases, the optimal velocity shows a transition from a finite value to zero, which means that purely diffusive motions are more efficient in reducing the searching time. Depending upon the value of the absorption strength (reactivity) of the target, the transition of the optimal velocity becomes either continuous or discontinuous. In the parameter space spanned by diffusion constant and reactivity, we obtain the phase diagram separating a passive-efficient region from an active-efficient region and determine the point at which the continuous transition line meets the discontinuous line, corresponding to the tri-critical point of the phase transition. We also consider arbitrary initial conditions, e.g., with non-uniform spatial distribution and biased initial orientation, and discuss how the existence of the optimal velocity depends on the initial conditions.

This paper is organized as follows: In Sec II, we introduce the system considered
and derive the analytic solution of MST.
In Sec III, we discuss the transition behaviors of optimal velocity using the Landau approach. In Sec IV, the initial condition dependence of MST is presented in limiting cases of a vanishing and a large diffusion constants. 
Lastly, summary is given in Sec V.

\section{Model}

We consider an active particle modeled by a run-and-tumble particle (RTP) subject to thermal noises in a confined one-dimensional space, $x\in[-L,L]$.
The particle has a self-propulsion velocity $v$ which flips its direction with a rate $\gamma$ by the Poisson process, and 
its dynamics is described by the overdamped Langevin equation,
\begin{equation}
\dot{x}=\pm v + \xi
\end{equation}
where $\xi$ is the random force, related to the diffusion constant $D$ via $\langle \xi (t) \xi (t') \rangle = 2 D \delta (t - t')$.
The target is located at the origin and in general, the interaction energy with the target is finite, leading to a finite absorption (reaction) strength.
The presence of a partially absorbing target is incorporated in the Fokker-Planck equation 
by a sink term given as a delta function with a strength $k$:
\begin{equation}\label{fp1d}
    \frac{\partial P_\pm}{\partial t}=D\frac{\partial^2 P_\pm}{\partial x^2}\mp v\frac{\partial P_\pm}{\partial x}+\gamma(P_\mp-P_\pm)-k \delta(x) \, P_\pm
\end{equation}
where $P_{\pm}(x,t|x_0)$ describes the probability distribution function of the particle, initially started at $x_0$, which has right or left propulsion direction.
 $k$ indicates the absorption strength (or reactivity) of the target, and the limit of $k \rightarrow \infty$ represents an ideal target which
perfectly absorbs the particle upon encountering.
The boundary condition at the target location is obtained by integrating Eq.~\eqref{fp1d} over $x\in[-\epsilon, \epsilon]$ with $\epsilon \rightarrow 0$, which 
leads to the so-called Robin (radiation) boundary condition:
$J_\pm (0^{-}) - J_\pm (0^+) = k P_\pm (0)$
where $J_\pm \equiv -D\partial_xP_\pm\pm vP_\pm$ is the probability current.
Here, an uniform initial distribution $P_\pm(x_0)=1/2L$ is considered.
Due to the symmetry of the system, $P_\pm ( x, t) = P_\mp (-x, t)$, and thus, it suffices only to consider the half domain, i.e., $x \in [0, L]$.
Assuming a hard wall at the domain boundary, we have a vanishing current, $J_\pm (L, t) =0$.
Using the symmetry, the boundary condition at the target location can be written as
\begin{equation}
\label{x=L}
- J_\mp (0^+, t) - J_\pm (0^+, t) = k P_\pm (0, t) ,
\end{equation}
and $P_+(0,t)=P_-(0,t)$.
Integrating Eq.~\eqref{fp1d} with respect to the time and initial distribution, we obtain
\begin{equation}
\label{int_fp1d}
    -\frac{1}{2L}=D\frac{\partial^2 \phi_\pm}{\partial x^2}\mp v\frac{\partial \phi_\pm}{\partial x}+\gamma(\phi_\mp-\phi_\pm)
\end{equation}
where $\phi_\pm$ is the time-integrated probability distribution,
\begin{equation}
    \phi_\pm(x)=\int^L_0 \, dx_0 \, P_\pm(x_0)\int^\infty_0 dt P_\pm(x,t|x_0) .
\end{equation}

To simplify further, we introduce $f=\phi_++\phi_-$ and $g=\phi_+-\phi_-$, and then Eq.~\eqref{int_fp1d} becomes
\begin{align}
    -\frac{1}{L}&=D \frac{\partial^2 f}{\partial x^2} -v \frac{\partial g}{\partial x} \label{int_fp1}\\
    0&=D  \frac{\partial^2 g}{\partial x^2}-v \frac{\partial f}{\partial x} -2\gamma g\label{int_fp2} .
\end{align}
The boundary conditions for $f$ and $g$ are now written as
$ -D \partial_x f (L) + v g(L) = 0,  -D \partial_x g(L) + v f(L) = 0, - D \partial_x f (0) + k f(0)/2 =0,$ and $ g(0) = 0$.
Integrating Eqs.~\eqref{int_fp1} and \eqref{int_fp2} under these boundary conditions, we obtain
\begin{align}
f(x)&=\frac{vA}{wD}\cosh{wx}+\frac{vB}{wD}\sinh{wx}-\frac{\gamma (x-L)^2}{w^2 LD^2}+C , \nonumber \\
g(x)&=A\sinh{wx}+B\cosh{wx}+\frac{v(x-L)}{w^2LD^2} ~,
\end{align}
where $A, B,$ and $C$ read as
\begin{equation}
    A=\frac{vwD}{v^2+2\gamma D\cosh wL}\left(\frac{2}{k}+\frac{\gamma L^2-D}{w^2 D^2 L}-\frac{2\gamma}{w^3 D^2}\sinh wL\right) , \nonumber
\end{equation}   
$ B=v/w^2D^2 , $ and 
$ C =2/k+\gamma L/w^2 D^2-vA/wD$
with $w^2=(v^2+2\gamma D)/D^2$.
Then, the average time for the particle to be absorbed by the target, referred to as the mean searching time (MST), is given by
\begin{equation}
    \left\langle t \right\rangle=\int^L_0 dx\ f(x) .
\end{equation}
After integration, the rescaled MST is expressed as
\begin{equation}
\label{mfpt}
  \MST =\frac{3\tilde{D}-1}{3\varphi^2}+\frac{\tilde{D}}{\tilde{v}^2\sech\Omega+2\tilde{D}}\left(2H+\frac{\tilde{v}^2}{\varphi}S\right)
\end{equation}
where we introduce the dimensionless parameters: $\tilde{v}=v /\gamma L, \tilde{D}= D/\gamma L^2, \tilde{k}= k/\gamma L, \varphi=\sqrt{\tilde{v}^2+2\tilde{D}}$ and $\Omega=\varphi/\tilde{D}$.
We rescale times by the inverse of the tumbling rate, according to $\tilde{t} = t \gamma$.
Here, the constants $H$ and $S$ are 
\begin{align}
    H&=\frac{2}{\tilde{k}}+\frac{1-\tilde{D}}{\varphi^2}+\frac{\tilde{v}^2}{\varphi^3}\tanh\Omega\\
    S&=\left(\frac{2}{\tilde{k}}+\frac{1-\tilde{D}}{\varphi^2}\right)\tanh\Omega+\frac{\tilde{v}^2-2\tilde{D}}{\varphi^3}(1-\sech\Omega) . \nonumber
\end{align}
This is one of our main results, the analytic expression of MST for a reactive target with an arbitrary absorption strength by an 1d RTP in the presence of the diffusion. 

\begin{figure}
\centering
\includegraphics[width = 0.45\textwidth]{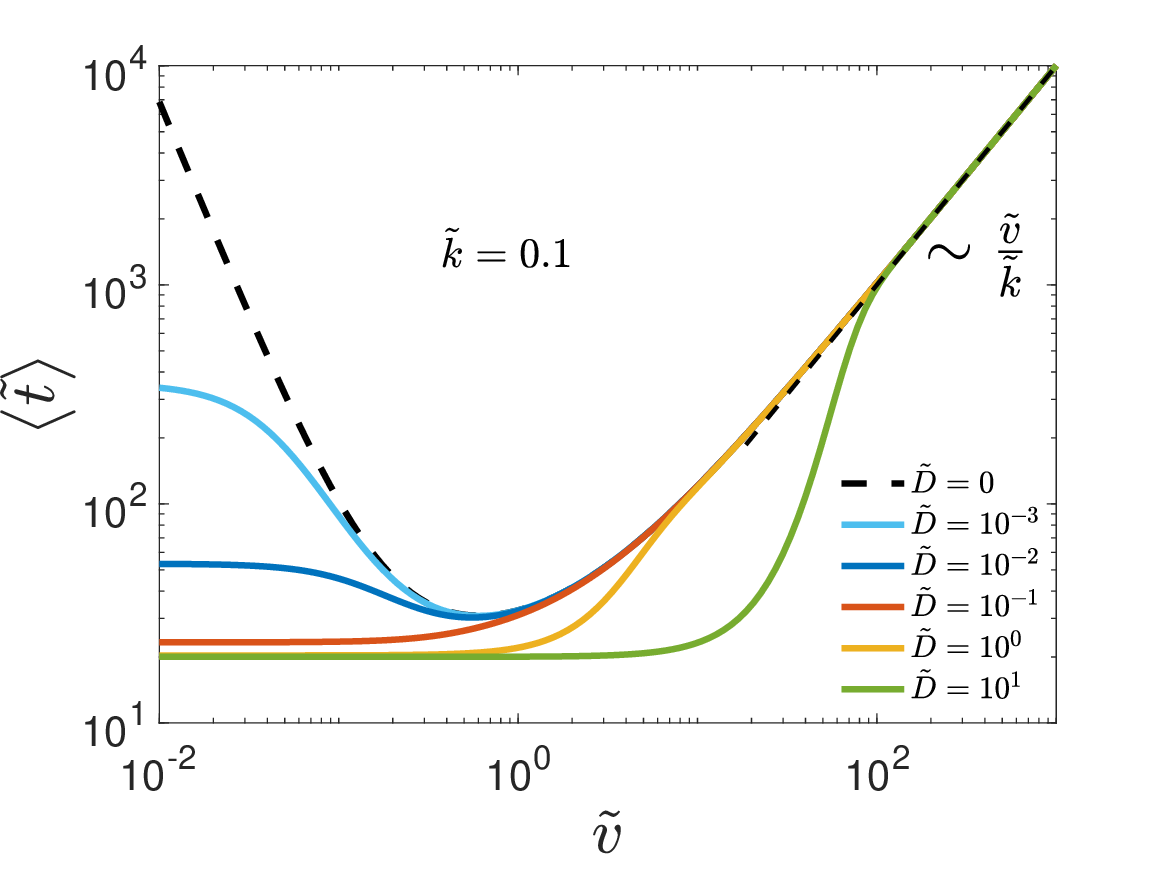}
\caption{\label{mfpt_1d} 
The mean searching time (MST) as a function of a dimensionless self-propulsion velocity $\tilde{v}$, which is evaluated using Eq.~(\ref{mfpt}) for various diffusion constants at an absorption strength, $\tilde{k}=0.1$. For small $\tilde{v} \ll 1$, MST shows plateaus of different heights, depending on $\tilde{D}$. As $\tilde{v}$ increases, MSTs with different $\tilde{D}$ converge on a dashed line, corresponding to MST of an RTP with $\tilde{D}=0$ given by Eq.~\eqref{time_D0}.}
\end{figure}

In Fig.\ref{mfpt_1d}, we present MST, $\MST$ (in units of an inverse of the tumbling rate $\gamma$), using Eq.(\ref{mfpt}) as a function of $\tilde{v}$ for various values of $\tilde{D}$ when $\tilde{k} =0.1$. The dashed line denotes the limiting case of $D \rightarrow 0$, i.e., an RTP in the absence of diffusion.
It is found that there exists a crossover velocity $\tilde{v}^\times$ from a diffusion-dominated regime to an activeness-dominated regime:
When $\tilde{v}$ is smaller than $\tilde{v}^\times$, the diffusion dominates and $\MST$ does not depend on $\tilde{v}$, leading to plateaus in Fig.~\ref{mfpt_1d}.
On the other hand, when $\tilde{v} \gg \tilde{v}^\times$, the diffusion is negligible and $\MST$ merges with that of an RTP with $D \to 0$ (dashed line).

The crossover velocity $\tilde{v}^\times$ can be determined by considering the diffusive limit of $\tilde{v} \ll 1$ or the ballistic limit of $\tilde{v} \gg 1$, separately.
In the diffusive limit where $\tilde{v} \ll 1$, the traveling distance of the particle by the propulsion velocity $v$ in a single run mode, i.e., between successive tumbling events, is much smaller than the system size $L$, and the trajectories of the particle look like diffusive motions with an effective diffusion constant of $\tilde{D}_{\text{eff}}=\tilde{D}+\tilde{v}^2/2$~\cite{Bech16}.
Therefore, if $\tilde{v}\ll \tilde{v}^\times = \sqrt{\tilde{D}}$, it is indistinguishable from a purely diffusive motion with a diffusion constant $\tilde{D}$, and the searching time is simply given by that of a passive Brownian particle \cite{Vasquez21},
\begin{equation}
\label{time_v0}
    \MST \simeq \frac{1}{3\tilde{D}_\text{eff}}+\frac{2}{\tilde{k}} \simeq \frac{1}{3\tilde{D}}+\frac{2}{\tilde{k}} ,
\end{equation}
which is independent of $\tilde{v}$.
If $\tilde{v}^\times=\sqrt{\tilde{D}} \ll \tilde{v}\ll 1$, the motion is still diffusive but with a substantial enhanced effective diffusion constant $\tilde{D}_{\text{eff}} \approx \tilde{v}^2/2$, which leads to 
\begin{equation}
\label{time_v0_2}
    \MST \simeq\frac{2}{3\tilde{v}^2}+\frac{2}{\tilde{k}} .
\end{equation}

In the ballistic limit where $\tilde{v} \gg 1$, the particle travels a distance larger than $L$ in a single run mode of the time scale of $\gamma^{-1}$.
The interpretation through the effective diffusion is thus no longer valid.
Instead, in a single run mode, the particle behaves like being subject to a constant force in one direction,
and the probability distribution follows the Boltzmann distribution, $P_\pm \propto \exp ({\pm v x/D}) = e^{\mathrm{Pe} \,  \tilde{x}}$,
with the P\'{e}clet number $\mathrm{Pe}=\tilde{v}/\tilde{D}$~\cite{Redner2001}.
If $1\ll\tilde{v}\ll \tilde{v}^\times = \tilde{D}$, the effective potential height generated by the propulsion velocity is much smaller than the thermal energy,
which renders the diffusion dominated. As a result, MST is again given by that of a passive particle, Eq (\ref{time_v0}).
If $1\ll \tilde{v}^\times = \tilde{D} \ll\tilde{v}$, the thermal energy is negligible compared to the effective potential strength and the probability distribution is highly concentrated 
near a confining boundary. Then, dynamics is well approximated by an RTP in the absence of diffusion.
In the limit of $D \to 0$, MST by an RTP can be readily found as
\begin{equation}
\label{time_D0}
    \MST \bigg|_{D\to0}=\frac{2}{3\tilde{v}^2}+\frac{3}{2\tilde{v}}+\frac{1}{2}+\frac{\tilde{v}+2}{\tilde{k}} ,
\end{equation}
which is plotted as a dashed line in Fig.~\ref{mfpt_1d} and reproduces Eq.~\eqref{time_v0_2} when $\tilde{v}\ll1$.
What is shown in Fig.~\ref{mfpt_1d} is consistent with the above argument.
For small $\tilde{v}$, MSTs show plateaus of different heights given as Eq.~(\ref{time_v0}), but for larger $\tilde{v}$, they converge toward that of an RTP, Eq.~\eqref{time_D0}.
In the diffusive limit, $\tilde{v}^\times \sim \sqrt{\tilde{D}}$ for $\tilde{D}\ll 1$,
while in the ballistic limit, $\tilde{v}^\times \sim \tilde{D}$ for $\tilde{D}\gg 1$.


\section{Transitions of optimal velocity}
\label{sec201}

\begin{figure}[t]
\centering
\includegraphics[width = 0.45\textwidth]{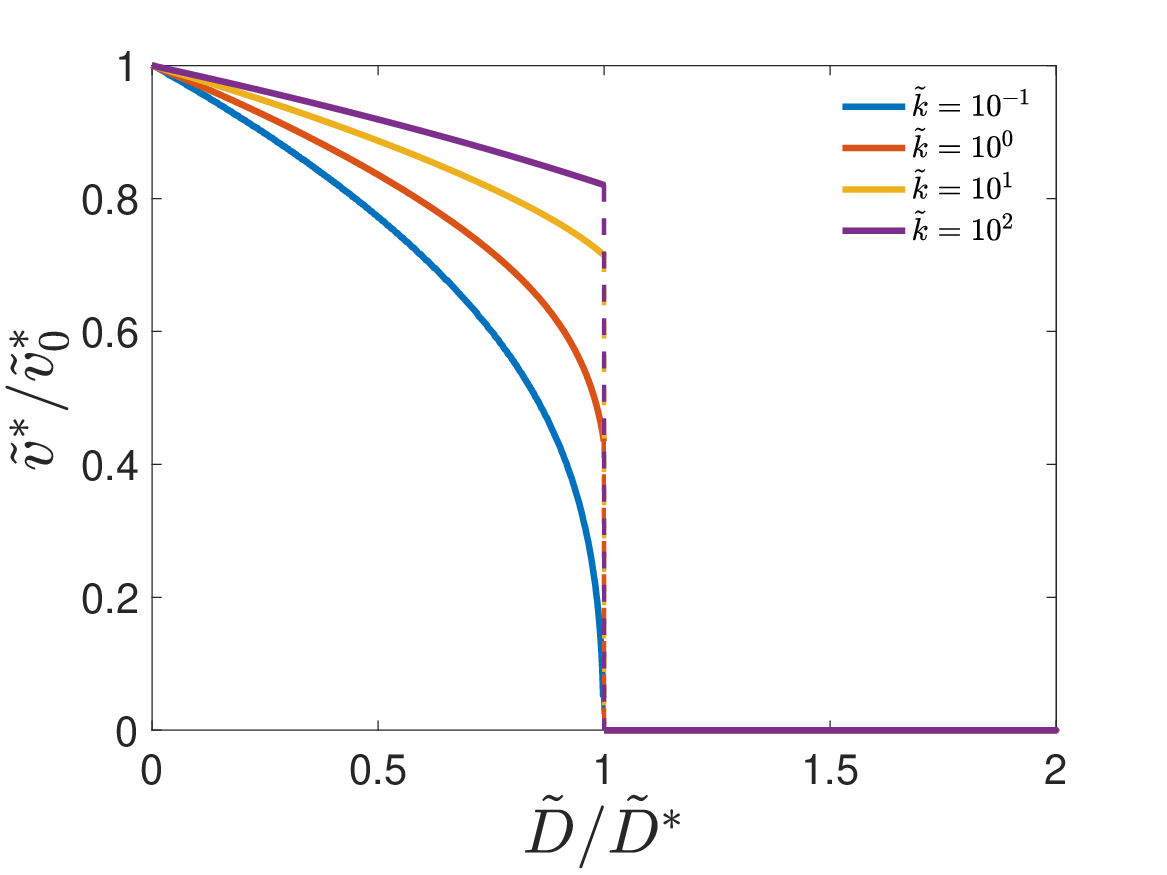}
\caption{\label{trans} The optimal self-propulsion velocity $\tilde{v}^*$ as a function of a dimensionless diffusion constant $\tilde{D}$. $\tilde{v}^\ast$ is numerically determined from Eq.~(\ref{mfpt}) for given $\tilde{D}$ and $\tilde{k}$. The optimal velocity $\tilde{v}^\ast$ continuously changes from a finite value to zero for small $\tilde{k}$,
while it changes discontinuously for large $\tilde{k}$. $\tilde{v}^*_0$ and $\tilde{D}^*$ are also numerically determined (see main text).
}
\end{figure}

Figure \ref{mfpt_1d} exhibits the existence of the optimal velocity $\tilde{v}^*$ at which MST, $\MST$, is minimized for given absorption strength (reactivity) $\tilde{k}$ and diffusion constant $\tilde{D}$.
It is also found that for small $\tilde{k}$,  $\tilde{v}^*$ continuously changes from a finite value to zero ({\it continuous transition}),
while for large $\tilde{k}$,  $\tilde{v}^*$ discontinuously jumps to zero ({\it discontinuous transition}) as $\tilde{D}$ increases.
In Fig. \ref{trans}, we plot the optimal velocity $\tilde{v}^*$ as a function of $\tilde{D}$ for various values of $\tilde{k}$, 
which clearly shows that there exists a transition point $\tilde{D}^*$ above which the optimal velocity $\tilde{v}^*$ becomes zero
and a purely diffusive motion performs better in terms of the target search (a passive-efficient state).
Depending on the absorption strength $\tilde{k}$, the transition from an active-efficient state, where MST is minimum for a particle with a finite self-propulsion velocity, to a passive-efficient state is continuous or discontinuous.

\begin{figure} [t]
\centering
\includegraphics[width = 0.45\textwidth]{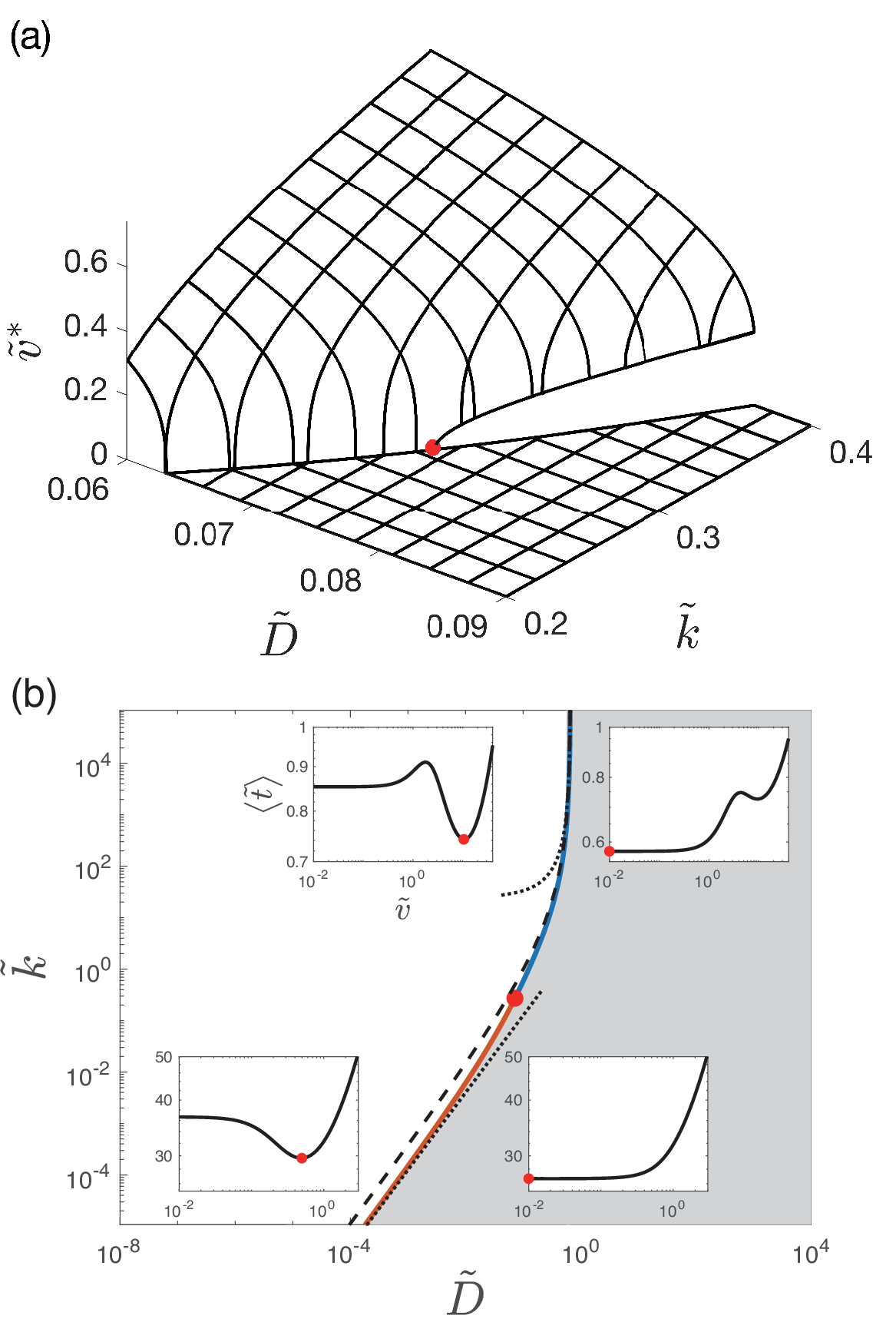}
\caption{\label{3d} (a) Optimal self-propulsion velocity $\tilde{v}^*$ as a function of $\tilde{D}$ and $\tilde{k}$. 
For small $\tilde{k}$, $\tilde{v}^*$ continuously changes from a finite value to zero as increasing $\tilde{D}$,
while for large $\tilde{k}$, it discontinuously changes.
The red dot indicates a point of $(\tilde{D}^\ast_c, \tilde{k}^\ast_c)\simeq(0.074,0.286)$, corresponding to the tri-critical point of Landau theory.
(b) Phase diagram indicating a passive-efficient state (gray region) and an active-efficient state (white region).
The transition line separating two regions is a solid blue(red) line for a discontinuous(continuous) transition.
Dashed line is obtained by solving Eqs.~\eqref{D_k} and \eqref{op_v}.
Dotted lines are asymptotes for $\tilde{k} \ll 1$ and $\tilde{k} \gg 1$, respectively, given by \eqref{asymp1} and \eqref{asymp2}.
Insets exhibit typical behaviors of $\MST$ vs. $\tilde{v}$ for corresponding parameters, where red dots indicate optimal points.
}
\end{figure}

In order to grasp a comprehensive picture, we adopt the Landau approach.
Suppose that near the transition point, $\MST$ is assumed to be analytic and is expanded as a polynomial function of $\tilde{v}$:
\begin{equation}
    \MST \approx t_0 +\frac{a}{2!}\tilde{v}^2+\frac{b}{4!}\tilde{v}^4+\frac{c}{6!}\tilde{v}^6+O(\tilde{v}^8)
\end{equation}
where $a (\tilde{D}, \tilde{k}) =\partial^2_{\tilde{v}} \MST|_{v=0}, b (\tilde{D}, \tilde{k}) = \partial^4_{\tilde{v}} \MST|_{v=0}$ and $c ( \tilde{D}, \tilde{k}) =\partial^6_{\tilde{v}} \MST|_{v=0}$.
Due to the symmetry of the system with respect to $v \rightarrow -v$, $\MST$ does not contain terms of odd order in $\tilde{v}$.
According to the Landau theory, when the free energy $F[m]$ is given as a polynomial of degree six as above, the order parameter $m^\ast$ that minimizes $F[m]$ shows a continuous phase transition for $b > 0$ or a discontinuous phase transition for $b < 0$~\cite{Huang}.
The point of $(a,b) = (0,0)$ corresponds to the tri-critical point \cite{Huang,plischke1994}.
In Fig.~\ref{3d}(a), we present the optimal velocity $\tilde{v}^*$, numerically determined using Eq.~(\ref{mfpt}), as a function of $\tilde{D}$ and $\tilde{k}$.
Clearly, for a given $\tilde{k}$, the optimal velocity $\tilde{v}^*$ is finite for $\tilde{D} < \tilde{D}^\ast$, while $\tilde{v}^*=0$ for $\tilde{D} > \tilde{D}^\ast$.
When $\tilde{k} < \tilde{k}^\ast$, change of $\tilde{v}^*$ is continuous as increasing $\tilde{D}$.
When $\tilde{k} > \tilde{k}^\ast$, the transition becomes discontinuous.
What we found is summarized in Fig.~\ref{3d}(b) where the phase diagram is depicted in the parameter plane of $\tilde{D}$ and $\tilde{k}$.
Insets display typical behaviors of $\MST$ as a function of $\tilde{v}$ for corresponding parameter regions.
In Fig.~\ref{3d}(b), the passive-efficient state (where a Brownian particle outperforms an active particle in searching time) is denoted as a gray region, and the active-efficient state (where a RTP with a finite propulsion velocity performs better) is as a white region.
The transition line $\tilde{D}^\ast (\tilde{k})$ separating two regions is indicated by a blue line for a discontinuous transition or by a red line for a continuous transition.
The discontinuous transition line meets with the continuous transition line at a point $(\tilde{D}^\ast_c, \tilde{k}^\ast_c)$, which corresponds to the tri-critical point of the Landau theory.
It is numerically found that $(\tilde{D}^\ast_c, \tilde{k}^\ast_c) \simeq (0.074, 0.285) \pm (0.001,0.001)$.

How can we determine the transition line $\tilde{D}^\ast (\tilde{k})$?
Full analytic expression of $\tilde{D}^\ast (\tilde{k})$ is hard to obtain, but
the asymptotic behaviors can be investigated as follows.
It is suggested in Fig.~\ref{mfpt_1d} that when $\tilde{D}$ is small, the minimum searching time is given by that of an RTP without diffusion.
At elevated $\tilde{D} > \tilde{D}^\ast$, the minimum searching time is now given by $\MST$ of a diffusive particle with $\tilde{v} =0$. 
Thus, we determine $\tilde{D}^\ast (\tilde{k})$ by comparing two searching times, i.e., 
the minimum searching time of an RTP with $\tilde{D} \to 0$ and the searching time of a purely diffusive particle with $\tilde{v}=0$. 
For a diffusive particle of $\tilde{v}=0$, $\MST$ is given by Eq.~\eqref{time_v0}, and
for $\tilde{D} \to 0$, $\MST$ is given by Eq.~\eqref{time_D0}.
Therefore, the transition line can be obtained by solving the following equation for $\tilde{D}^\ast (\tilde{k})$:
\begin{equation}
\label{D_k}
    \frac{1}{3\tilde{D}^*}+\frac{2}{\tilde{k}} = \frac{2}{3\tilde{v}^{*2}_0}+\frac{3}{2\tilde{v}^*_0}+\frac{1}{2}+\frac{\tilde{v}^*_0+2}{\tilde{k}}
\end{equation}
where $\tilde{v}^*_0(\tilde{k})$ is the optimal velocity in the limit of $D \to 0$, which is determined from the derivative of Eq.~\eqref{time_D0},
\begin{equation}
\label{op_v}
    \partial_{\tilde{v}} \MST \bigg|_{D\to0}=-\frac{4}{3\tilde{v}^{*3}_0}-\frac{3}{2\tilde{v}^{*2}_0}+\frac{1}{\tilde{k}} = 0.
\end{equation}
For $\tilde{k}\ll1$,
   $ \tilde{v}_0^* \simeq (4/3)^{1/3}\tilde{k}^{1/3} $
and for $\tilde{k}\gg1$,
$ \tilde{v}_0^*\simeq (3/2)^{1/2}\tilde{k}^{1/2} $.
Inserting these values of $ \tilde{v}_0^*$ into Eq.~\eqref{D_k}, we find the asymptotic behaviors of $\tilde{D}^\ast (\tilde{k})$:
For $\tilde{k}\ll1$,
\begin{equation}
\label{asymp1}
     \tilde{D}^*(\tilde{k})\sim \tilde{k}^{2/3}
\end{equation}
and for $\tilde{k}\gg1$,
\begin{equation}
\label{asymp2}
    \tilde{D}^*(\tilde{k})\simeq\frac{2}{3}\left(1-\frac{2\sqrt{6}}{\sqrt{\tilde{k}}}\right) .
\end{equation}
In Fig.~\ref{3d}(b), the transition line $\tilde{D}^*(\tilde{k})$ obtained by solving Eq.~\eqref{D_k} combined with Eq.~\eqref{op_v} is shown
as a dashed line.
The asymptotic behaviors, Eqs.~\eqref{asymp1} and \eqref{asymp2}, are shown as dotted lines,
which show a reasonable agreement with numerically obtained transition line, respectively, in the limits of $\tilde{k} \ll 1$
and $\tilde{k} \gg 1$.
Equation \eqref{asymp2} also indicates the existence of an upper bound $D^\ast_{upper}$ of $\tilde{D}^*(\tilde{k})$.
When $\tilde{D} > \tilde{D}^\ast_{upper} = 2/3 $,
a purely passive Brownian motion is always an optimal solution for {\it any value of} $k$ that minimizes the searching time.
For example, consider the limit of $k \to \infty$, i.e., an ideal target with a perfect absorption. 
For $\tilde{D} < \tilde{D}^\ast_{upper}$, we find that $v^\ast \to \infty$ and $\langle t \rangle \to 1/2 \gamma$.
For $\tilde{D} > \tilde{D}^\ast_{upper}$, $v^\ast $ suddenly changes to zero and $\langle t \rangle$ is simply given by $L^2/3D$.
Equating two searching times, $L^2/3 D^\ast_{upper} = 1/ 2 \gamma$, one retrieves $\tilde{D}^\ast_{upper} = 2/3$.

\section{Initial condition dependence on Transitional behavior}
\label{sec202}

So far, we have considered a uniform spatial distribution and a random orientation of a particle as an initial condition.
As different initial conditions can be regarded, it would be of interest to investigate how the existence of the optimal velocity does depend on a given initial condition.
For this purpose, we consider an arbitrary initial distribution of position and orientation of propulsion direction:
$p(x_0)$ is the initial spatial distribution on the domain $x_0 \in[-L,L]$,
and $P_{R (L)} (x_0)$ is the probability of an initial orientation at $x=x_0$, i.e., initially heading to the right (left).
Obviously, $P_R (x_0) +P_L (x_0) =1$.
For arbitrary initial distributions, it is difficult to solve the Fokker-Planck equation, Eq.~\eqref{fp1d}.
Our strategy to tackle this problem is to decompose MST obtained in the previous sections into two components:
\begin{equation}
\label{total_t}
    \MST=\MST_i+\MST_a .
\end{equation}
Here, $\MST_i$ is the mean first-passage time (MFPT), i.e., the time it takes for a particle to arrive at the target {\it for the first time},
and $\MST_a$ is the mean absorption time (MAT), i.e., the time it takes for a particle arrived at the target to be finally absorbed.
For a perfectly absorbing target ($k \to \infty$), the particle is instantaneously absorbed, leading to $\MST_a \to 0$ and $\MST \rightarrow \MST_i$,
but for a target with a finite $k$, MAT can be obtained from Eq.~(\ref{mfpt}) by collecting terms which depends on the absorption strength $\tilde{k}$:
\begin{equation}
\label{ab_time}
    \MST_a = \left( \frac{2}{\tilde{k}} \right) \frac{1}{\sech{\Omega}+2\tilde{D}/\tilde{v}^2}  \left(\frac{2\tilde{D}}{\tilde{v}^2}+\frac{\tanh{\Omega}}{\Omega}\right) .
\end{equation}
For a given initial distribution, MFPT $\MST_i$ is obtained by solving Eq.~\eqref{fp1d} under a perfect absorbing boundary,
i.e., taking the limit of $k \to \infty$.
The advantage of this decomposition is that all the initial distribution dependence is isolated in MFPT, $\MST_i$, and
MAT, $\MST_a$ does not depend on the initial distribution.
Note that MAT, $\MST_a$, is identical to the mean searching time when the particle initially starts from the target position.
As a result, our task to find MST, $\MST$, for an arbitrary initial distribution reduces to calculation of the MFPT $\MST_i$ for the given distribution.
Then, adding $\MST_a$ given by Eq.~\eqref{ab_time}, MST $\MST$ is evaluated as in Eq.~\eqref{total_t}.
However, it is still non-trivial to obtain $\MST_i$ for an arbitrary initial distribution. 
In the following, we only consider two limiting cases where the analytic expression of $\MST_i$ can be found.

\subsection{$D \to 0$ case}
First, we consider an RTP in the absence of diffusion. Let the particle to be initially injected on the right half of the domain, $x\in(0,L]$. 
To find $\MST_i$, we set a perfect absorption boundary at $x=0$. When $D=0$, Eq.~\eqref{fp1d} reduces to
\begin{equation}
    \frac{\partial P_\pm}{\partial t}=\mp v\frac{\partial P_\pm}{\partial x}+\gamma(P_\mp-P_\pm) .
\end{equation}
Similar to Sec II, we introduce the time-integrated probability distribution (but not with respect to the initial position), 
$\phi_\pm=\int_0^\infty P_\pm(x,t|x_0)dt$.
Defining $f=\phi_++\phi_-$ and $g=\phi_+-\phi_-$, one has
\begin{align}
    -\delta(x-x_0)&=-v \, \partial_x g  \nonumber \\
    -\eta (x_0)\, \delta(x-x_0)&=-v\,  \partial_x f -2\gamma g \label{fp_b} .
\end{align}
Here, $\eta (x_0)=P_R(x_0)-P_L(x_0)$ indicates the degree of a bias in the initial orientation. Let $N_+(t)$ to be the concentration on the confining hard wall at $x=L$, then it satisfies the following equations \cite{jeon23}.
\begin{align}
    \partial_tN_+&=vP_+(x=L)-\gamma N_+ \nonumber \\
    0&=vP_-(x=L)-\gamma N_+ .
\end{align}
Denoting $W_+=\int_0^\infty N_+(t)dt$, one shows that 
\begin{equation}
\label{Wp}
    W_+=\frac{v}{\gamma}p_+(x=L)=\frac{v}{\gamma}p_-(x=L)
\end{equation}
where $p_\pm(x=L)=\int_0^\infty P_\pm(x=L)dt$. Solving Eqs.~\eqref{fp_b} under given boundary conditions, we find MFPT for a particle initially located on the right half domain, i.e., $x_0 \in(0,L]$, which is denoted as $\MST_{i}^r$:
\begin{eqnarray}
\label{fpt_d0+}
    \MST_{i}^r (\tilde{x}_0) &=& \int^1_0 \, f(\tilde{x}) d\tilde{x} + \gamma W_+  \nonumber \\
    &=&\frac{1}{\tilde{v}^2}\left(1-(1-\tilde{x}_0)^2\right) +\frac{1+ \eta(\tilde{x}_0) }{2}  \nonumber \\
    &&+\frac{1}{\tilde{v}} \bigg( \big(1+ \eta(\tilde{x}_0)) +\tilde{x}_0(1-\eta(\tilde{x}_0) \big)\bigg) 
\end{eqnarray}
where $\tilde{x}_0=x_0/L$ and $\tilde{t}=t\gamma$. 
Note that if we set $x_0=0$ with $P_R(x_0)=1$, Eq.~\eqref{fpt_d0+} reproduces the mean returning time of an RTP given as $\MST_i=1+2/\tilde{v}$~\cite{jeon23}.

Let us now consider the situation where particle is initially injected to the opposite half domain, $x_0\in[-L,0)$, with $\eta(x_0)= P_R(x_0)-P_L(x_0)$. 
Since we have a spatial inversion symmetry, the equation of motion remains the same as we flip the whole system. 
Consequently, $\MST_{i}^\ell$ can be obtained from $\MST_{i}^r$ with $x_0\to-x_0$ and $\eta\to -\eta$.
Taking the limit of $\tilde{D} \to 0$ in Eq.~\eqref{ab_time}, MAT is given as
\begin{equation}
\MST_a = \frac{\tilde{v} + 2}{\tilde{k}} ,
\end{equation}
which recovers the $\tilde{k}$-dependent term of Eq.~\eqref{time_D0}~\cite{note}.
For a given initial distribution $p(x_0)$, MST then reads 
\begin{eqnarray}
    \MST &=&\int^{1}_{-1}\MST_i (\tilde{x}_0)\, p(\tilde{x}_0)d\tilde{x}_0 + \MST_a \nonumber \\
    &=& \int^{1}_{0} \MST_{i}^r \, p(\tilde{x}_0) d\tilde{x}_0 + \int^{0}_{-1} \MST_{i}^\ell \, p(\tilde{x}_0) d\tilde{x}_0 + \frac{\tilde{v}+2}{\tilde{k} } \nonumber \\
   &=&\frac{A_0}{\tilde{v}^2}+\frac{2B_0}{\tilde{v}}+C_0+\frac{\tilde{v}+2}{\tilde{k}}  \label{fpt_d0}
\end{eqnarray}
where
\begin{align}
    A_0&=\int^1_{-1}w_1(\tilde{x}_0)p(\tilde{x}_0)d\tilde{x}_0\label{A} \nonumber \\
    B_0&=\int^1_{-1}w_2(\tilde{x}_0)p(\tilde{x}_0)d\tilde{x}_0\nonumber \\
    C_0&=\frac{1}{2}\int^1_{-1} \eta(\tilde{x}_0) p(\tilde{x}_0)d\tilde{x}_0+\frac{1}{2} .
\end{align}
The weighting functions $w_1(\tilde{x}_0)$ and $w_2(\tilde{x}_0)$ are given as 
\begin{align}
    w_1(\tilde{x}_0)&=1-(1-|\tilde{x}_0|)^2 \nonumber \\
    w_2(\tilde{x}_0)&=|P_L-\theta(\tilde{x}_0)|+|P_R-\theta(\tilde{x}_0)||\tilde{x}_0| .
\end{align}
Since the constants $A_0,B_0$ and $C_0$ are all positive, Eq.~\eqref{fpt_d0} clearly shows that for an RTP with $D=0$,
there always exists a finite optimal velocity $\tilde{v}^\ast$ that minimizes the searching time for any initial condition.
As a trivial check, one considers the uniform initial distributions, $p(\tilde{x}_0)=1/2$ and $P_R(x_0)=P_L(x_0)=1/2$, then Eq.~\eqref{fpt_d0} recovers the previous result, Eq.~\eqref{time_D0}.

\subsection{$D\gg1$ case}
Now we consider the opposite limit of a large diffusion constant:
When $\tilde{D} \gg 1$, or equivalently, $L^2/D \ll 1/\gamma$, the diffusion time scale on which the particle travels a distance of the system size via diffusion is much smaller than the flipping time scale. 
This means that before the next tumbling occurs, the particle arrives at the target position.
Thus, MFPT can be obtained by assuming a static situation where the self-propulsion direction does not flip.
For a particle initially injected on the right domain $x_0\in(0,L]$, 
MFPT satisfies the backward Fokker-Planck equation:
\begin{equation}
\label{t0}
    -1=D \frac{\partial^2 }{\partial x_0^2} \uMST_{i \pm}^{r} \mp v \frac{\partial }{\partial x_0} \uMST_{i \pm}^{r}
\end{equation}
where $\uMST_{i\pm}^{r}$ is MFPT for a particle with an initial orientation to the right (left).
Considering a perfect absorbing boundary at $x=0$ and a hard wall at $x=L$, we obtain
\begin{equation}
    \uMST_{i+}^r=\frac{x_0}{v}+\left(\frac{L}{v}-\frac{D}{v^2}\right)(e^{vx_0/D}-1) ,
\end{equation}
which is expanded up to the first order of $v$ as
\begin{equation}
    \uMST_{i+}^r=\frac{L^2-(L-x_0)^2}{2D}+\frac{(3L-x_0)x_0^2}{6D^2}v+O(v^2) .
\end{equation}
Adding $\uMST_{i-}^r$ for a particle initially headed toward the target, 
\begin{equation}
    \uMST_{i}^r=\frac{L^2-(L-x_0)^2}{2D}+\frac{ \eta(x_0) (3L-x_0)x_0^2}{6D^2}v+O(v^2)
\end{equation}
where $\eta(x_0)=P_R(x_0)-P_L(x_0)$. Similarly, if we consider a particle initially at the opposite side of the  domain $x_0\in[-L,0)$, we obtain $\uMST_{i}^\ell$ by changing $x_0\to-x_0$ and $\eta\to-\eta$:
\begin{equation}
    \uMST_{i}^\ell=\frac{L^2-(L+x_0)^2}{2D}-\frac{ \eta(x_0)(3L+x_0)x_0^2}{6D^2}v+O(v^2)
\end{equation}
Summing up, MPFT (in units of $\gamma^{-1}$) with an initial distribution $p(x_0)$ over the whole domain reads as
\begin{equation}
\label{ti_approxi}
    \MST_i =\frac{A_0}{2\tilde{D}}+\frac{B_1}{6\tilde{D}^2}\tilde{v}+O(\tilde{v}^2) ,
\end{equation}
which is expressed in terms of dimensionless variables.
Here, $A_0$ is the same as before, i.e., given by Eq.~(\ref{A}), and 
\begin{equation}
    B_1=\int^1_{-1}w_3(\tilde{x}_0) \eta(\tilde{x}_0) p(\tilde{x}_0)d\tilde{x}_0
\end{equation}
where the length is again rescaled as $\tilde{x} = x/ L$ and 
\begin{equation}
    w_3(\tilde{x}_0)=\tilde{x}_0|\tilde{x}_0|(3-|\tilde{x}_0|) .
\end{equation}
Unlike $B_0$ in Eq.~\eqref{A}, $B_1$ depends on both initial spatial and directional distributions and can be negative.
When $\tilde{D}$ is finite, MAT of Eq.~\eqref{ab_time} is expanded for small $\tilde{v}$ as
\begin{equation}
    \MST_a 
    \simeq\frac{2}{\tilde{k}}+\frac{\tilde{v}^2}{\tilde{k}}\left(\frac{1}{\sqrt{2\tilde{D}}}\tanh{\sqrt{\frac{2}{\tilde{D}}}}-\frac{1}{\tilde{D}}\sech{\sqrt{\frac{2}{\tilde{D}}}}\right)+O(\tilde{v}^4)
\end{equation}
As $\MST_a$ does not contain a $\tilde{v}$-linear term unless $\tilde{D}$ is zero,
the leading order dependence of MST $\MST$ comes from $\MST_i$ through $B_1$.
Thus, if $B_1 < 0$, $\MST$ is always a decreasing function of $\tilde{v}$ for small $\tilde{v}$ and there exists a finite optimal velocity $\tilde{v}^\ast$.
Given that $w_3(\tilde{x}_0)$ is an odd and monotonically increasing function on $\tilde{x}_0\in[-1,1]$, one of the simplest ways to make $B_1<0$ is to set $ \eta(\tilde{x}_0)>0$ on $\tilde{x}_0<0$ and $\eta(\tilde{x}_0)<0$ on $\tilde{x}_0>0$, which corresponds to the case where the initial orientation of a particle is biased toward the target.

\begin{figure}
\centering
\includegraphics[width = 0.45\textwidth]{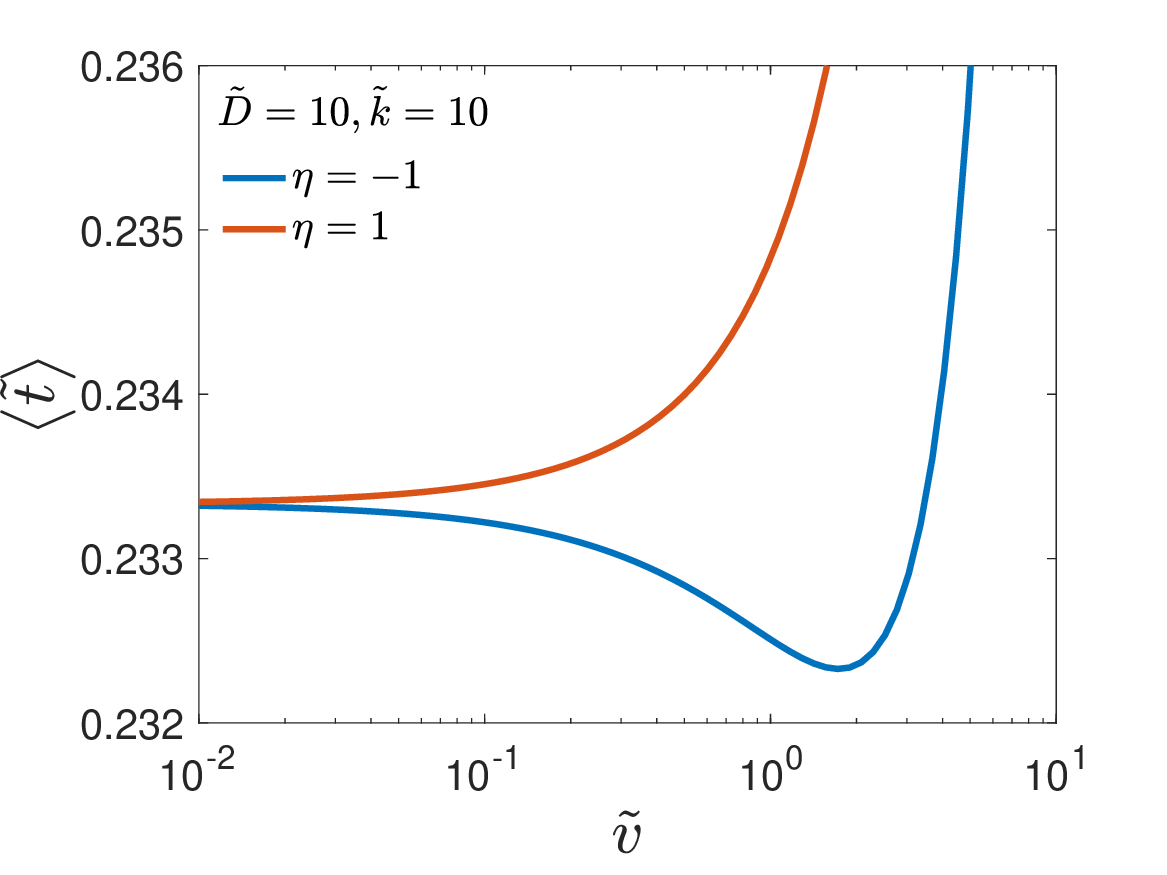}
\caption{\label{b-effect} MST $\MST$ as a function of $\tilde{v}$ for a system with initial uniform spatial distribution but biased in direction. $\MST$ is obtained as a sum of Eq.~(\ref{ab_time}) and (\ref{ti}). For $\eta>0$, MST increases monotonically, implying zero optimal speed, while for $\eta<0$, it shows non-monotonic behavior and has a finite optimal velocity even in the case of a large $\tilde{D}$.}
\end{figure}

\subsection{Uniform spatial distribution with $P_R\neq P_L$}

To confirm the results of previous section, we consider an example, i.e., a system of initially uniformly distributed in space $x\in(0,L]$ but biased in the orientation.
In this case, the full analytic expression of MST can be found.
Suppose that $p(x_0)=1/L$ and that the directional distribution is also a constant, $P_{R,L}(x_0)=P_{R,L}$, but with different values, $\eta=P_R-P_L\neq0$. If $\eta>0$, particle is initially biased toward the wall, and if $\eta<0$, the particle is initially biased toward the target. For a perfectly absorbing target at $x=0$ and a hard wall at $x=L$, $J_+(L,t)=J_-(L,t)=0$ and $P_+(0,t)=P_-(0,t)=0\label{x=0_2}$.
Solving the Fokker-Planck equation, Eq.~(\ref{fp1d}) under these boundary conditions, we obtain MFPT,
\begin{equation}
\label{ti}
    \MST_i=\frac{2}{3\varphi^2}+\frac{\eta\tilde{v}}{2\varphi^2}+\frac{1}{\tilde{v}^2\sech{\Omega}+2\tilde{D}}\cdot\frac{\tilde{v}\tilde{D}}{\varphi^2}\cdot(E+F)
\end{equation}
where
\begin{eqnarray}
    E&= &\left(\frac{\tilde{v}-\eta\tilde{D}}{\varphi^2}\right)\left\{(\tilde{v}^2-2\tilde{D})(1-\sech{\Omega})+2\varphi\tanh{\Omega}\right\} \nonumber \\
    F&= &\frac{\tilde{v}}{\varphi}(\eta\tilde{v}+1-\tilde{D})(\tanh{\Omega}-\Omega\sech{\Omega}) .
\end{eqnarray}
The MST is given by the sum of Eq.~(\ref{ab_time}) and (\ref{ti}), which is depicted in Fig.~\ref{b-effect} for positive (negative) values of $\eta$
for a system with $\tilde{D}=10$ and $\tilde{k}=10$. As expected, depending upon the sign of $\eta$, $\MST$ shows a monotonic or non-monotonic behavior.
$\MST$ can be expanded up to the first order of $\tilde{v}$ as
\begin{eqnarray}
    \MST &=&\frac{1}{3\tilde{D}}+\frac{2}{\tilde{k}}+\frac{\eta}{4\tilde{D}}\bigg[1+\tilde{D}\left(1-\sech{\sqrt{\frac{2}{\tilde{D}}}}\right) \nonumber \\
     &&-\sqrt{2\tilde{D}}\tanh{\sqrt{\frac{2}{\tilde{D}}}}\bigg]\tilde{v}+O(\tilde{v}^2) .
\end{eqnarray}
For $D\gg1$, $ \left\langle t \right\rangle$ is simplified into
\begin{equation}
	\MST \simeq\frac{1}{3\tilde{D}}+\frac{2}{\tilde{k}}+\frac{\eta}{8\tilde{D}^2}\tilde{v}+O(\tilde{v}^2) ,
\end{equation}
\\which agrees with the result of the previous section when the coefficients $A_0$ and $B_1$ of Eq.~(\ref{ti_approxi}) are evaluated using the given distributions.
When $\eta<0$, the searching time decreases with the propulsion velocity for $\tilde{v} \ll 1$. It means that the diffusive searching time is no longer efficient even if the diffusion constant is large.  This is intuitively understood by considering that $\eta<0$ implies the initial bias of the propulsion direction toward the target.


\section{Summary}
We have studied the target search problem of a reactive target by an active particle, represented by RTP, in the presence of thermal diffusion. Analytic expression of MST is found by solving the Fokker-Planck equation. 
For an uniform initial distribution, we show that there exists an optimal self-propulsion velocity $v^\ast$ at which the searching time is minimum.
When the reactivity of the target is weak, the optimal velocity continuously changes from a finite value to zero as the diffusion constant increases.
When the reactivity is strong, the optimal velocity exhibits a discontinuous transition.
Adopting the Landau picture, we explain the continuous (discontinuous) transition line in the parameter space of diffusion constant and reactivity.
Interestingly, there exists a regime where a simple diffusive particle performs better than an active particle in terms of searching time.
Finally, the dependence of the searching time on the initial condition is also investigated in the two limiting cases of a vanishing diffusion constant and a large diffusion constant.


\acknowledgements
This research was supported by the National Research Foundation of Korea (NRF) grant funded by the Korean government (MSIT) (RS-2023-00251561).


\begin{thebibliography}{99}
\bibitem{note}
When $\tilde{D} \to 0$, there are two vanishing length scales, i.e., the target size (the width of the delta function) and the length scale associated with the diffusion ($D/v$). Depending on which of the two length scales more rapidly converges zero, the mathematical expression of the boundary condition at the target position changes, which leads to slightly different expressions of MAT. This reflects the fact that if $\tilde{D} =0$, there is a discontinuity of the probability density at the target location.
\end{thebibliography}
\end{document}